\documentclass[%
 aip,
 jmp,%
 amsmath,amssymb,
 preprint,%
]{revtex4-1}

\usepackage{graphicx}
\usepackage{dcolumn}
\usepackage{bm}

\newcommand{\doubletA}{A_{{3}}\cos\,\theta\, +  A_{{4}}\sin\,\theta\,}

\newcommand{\doubletBB}{B_{{3}}\cos\,2\theta\, +  B_{{4}}\sin\,2\theta\,}
\newcommand{\doubletC}{C_{{1}}\sin\,3\theta\, +  C_{{2}}\cos\,3\theta\,}

\begin{document}


\title[{\footnotesize A.M. Escobar-Ruiz, J.C. L\'opez Vieyra, P.
Winternitz and \.{I}. Yurdu\c{s}en}]{Fourth-order
superintegrable systems separating in Polar Coordinates. II. Standard Potentials}

\author{Adrian M. Escobar-Ruiz}
\email{escobarr@crm.umontreal.ca}
\affiliation{
Centre de recherches math\'ematiques 
and D\'epartement de math\'ematiques \\
et de statistique, Universit\'e de Montreal,  C.P. 6128,
succ. Centre-ville,\\ Montr\'eal (QC) H3C 3J7, Canada}
\author{J. C. L\'opez Vieyra}%
\email{vieyra@nucleares.unam.mx}
\affiliation{ Instituto de Ciencias Nucleares, Universidad Nacional
Aut\'onoma de M\'exico, Apartado Postal 70-543, 04510 M\'exico, D.F., Mexico
}%

\author{P. Winternitz}
\email{wintern@crm.umontreal.ca}
\affiliation{
Centre de recherches math\'ematiques 
and D\'epartement de math\'ematiques \\
et de statistique, Universit\'e de Montreal,  C.P. 6128,
succ. Centre-ville,\\ Montr\'eal (QC) H3C 3J7, Canada}

\author{$\,\,\,\,\,\,\,\,\,\,\,\,\,\,\,\,\,\,\,\,\,\,\,\,\,\,\,\,\,\,\,\,$\.{I}. Yurdu\c{s}en}
\email{yurdusen@hacettepe.edu.tr}
\affiliation{
Centre de recherches math\'ematiques 
and D\'epartement de math\'ematiques \\
et de statistique, Universit\'e de Montreal,  C.P. 6128,
succ. Centre-ville,\\ Montr\'eal (QC) H3C 3J7, Canada}
\affiliation{Department of Mathematics, Hacettepe University,
                     06800 Beytepe,  \\ Ankara, Turkey}

\date{\today}

\begin{abstract}
Superintegrable Hamiltonian systems in a two-dimensional Euclidean
space are considered. We present all real \emph{standard} potentials that allow separation of variables
in polar coordinates and admit an independent fourth-order integral of
motion. The general form of the potentials satisfies a linear ODE.  In the classical case, the \emph{standard} potentials
coincide with the Tremblay-Turbiner-Winternitz (TTW) or Post-Winternitz (PW) models. In the quantum case new
superintegrable systems are obtained, in addition to the TTW and PW ones. Their classical limit is free motion.
\end{abstract}

\keywords{Superintegrability, separation of variables}
\maketitle

\section{INTRODUCTION}
The purpose of this paper is to obtain and classify all classical and quantum superintegrable Hamiltonians that allow separation of variables in polar coordinates and admit a fourth-order integral of motion $Y$. We focus on \emph{standard} potentials, namely those that satisfy
(not trivially) a linear ODE. This linear ODE is nothing but the compatibility condition for the existence of the integral $Y$. Superintegrable systems separating in polar coordinates, allowing fourth-order integrals of motion and involving
\emph{exotic} potentials were studied in an earlier article\cite{ELW}.

Roughly speaking, a Hamiltonian with $n$ degrees of freedom is called integrable if it allows $n$ independent well defined integrals of motion in involution. It is minimally superintegrable if it allows $n+1$ such integrals, and maximally superintegrable if it admits $2n-1$ integrals (where only subsets of $n$ integrals among them can be in involution).

The best known superintegrable systems are the harmonic oscillator with its $su(n+1)$ algebra of integrals, and the Kepler-Coulomb system with its $o(n+1)$ algebra (when restricted to fixed bound state energy values).

A more recent review article provides precise and detailed definitions, general settings and
motivation for studying superintegrable systems~\cite{MillerPostWinternitz:2013}.
A systematic search for superintegrable classical and quantum systems in $E_2$
and $E_3$ established an interesting connection between second-order superintegrability and
multiseparability in the Schr\"{o}dinger or Hamilton-Jacobi
equation~\cite{Capel:2015,Fris:1965,Fris:1966,Makarov:1967,KalninsMW:1976,Kalnins:2002,
Carinena:2017}.

An extensive literature exists on second-order superintegrability in spaces of
$2$, $3$ and $n$ dimensions, Riemannian and pseudo-Riemannian, real or complex,
\cite{Kalnins:2002, KalninsI:2005, KalninsII:2005, KalninsIII:2005, KalninsIV:2006,
KalninsV:2006}.

The systematic study of higher order integrability is more recent. Pioneering work
is due to Drach~\cite{Drach:1935,Drach:1935b}.  For more recent work
see~\cite{Bermudez:2016,Chanu:2011,Chanu:2012,Hietarinta:1998,Tsiganov:2000,GravelWinternitz:2002,Marquette:2009,
Marquette:2010, Ranada:2013,Celeghini:2013,Fernandez:2016,Hakobyan:2016,GungorKNN:2014,TTW:2009,TTW:2010,QuesneTTW:2010,
PostVinetZhedanov:2011,MarquetteSajediW:2017}.

The present article is a contribution to a
series~\cite{Abouamal:2018,GravelWinternitz:2002,Gravel:2004,TremblayW:2010,
PostWinternitz:2015,
MarchesielloPostSnobl:2015,PopperPW:2012,MarquetteSajediW:2017, ELW} devoted to
superintegrable systems in $E_2$ with one integral of order $n\ge 3$ and one of order $n\leq 2$. 

In this article we restrict ourselves to the space $E_2$. The Hamiltonian has
the form
\begin{equation}
\label{H0}
H\ =\ \frac{1}{2}(p_x^2+p_y^2) + V(x,y) \ ,
\end{equation}
$V(x,y)$ is a scalar potential and in classical mechanics $p_x$ and $p_y$ are the momenta conjugate to the
Cartesian coordinates $x$ and $y$. In quantum mechanics they are the
corresponding operators
$p_{x} = -i \hbar \frac{\partial}{\partial x}$, $p_{y} = -i \hbar
\frac{\partial}{\partial y}$. In {polar coordinates} $(x,y)\equiv
(r\cos\theta,\,r\sin\theta)$, the classical Hamiltonian reads
\begin{equation}
\label{H}
H\ =\ \frac{1}{2} \left(p_r^2 \, + \, \frac{p_\theta^2}{r^2}\right)  \ +  \  V(r,\theta)  \ ,
\end{equation}
where $p_r$ and $p_\theta$ are the associated canonical momenta. The corresponding quantum operator takes the form
\begin{equation}
\label{Hpolar}
H\ =\ -\frac{\hbar^2}{2}\,\bigg(\partial^2_r + \frac{1}{r}\partial_r \,  + \,  \frac{1}{r^2}\partial_\theta^2\bigg)  \ + \ V(r,\theta) \ .
\end{equation}

In paper \cite{ELW} we found all the \emph{exotic} potentials, i.e. those which do not satisfy any linear differential equation. Here we solve the complementary problem, namely we find all the \emph{standard} potentials which do satisfy a linear ODE. In all equations we keep the Planck constant $\hbar$ explicitly. Classical \emph{standard} potentials will be obtained in the limit $\hbar\rightarrow 0$.

Before going into the question of standard superintegrable potentials with a fourth-order integral of motion,
let us recall that two families of standard superintegrable potentials in $E_2$ are already known. Both allow
separation of variables in polar coordinates and hence admit the second-order integral
\begin{equation}
\label{X}
 X \ =\ p_\theta^2 \ + \ 2\,S(\theta)\ ,
 \qquad
p_{\theta} = L_z = -i \hbar \frac{\partial}{\partial \theta}\,.
\end{equation}
They also admit a further polynomial integral of arbitrary order $N$.
One of them is the TTW potential
\begin{eqnarray}
V_{\rm TTW} = b r^2 + \frac{1}{r^2} \left[\frac{\alpha}{\cos^2 (k \theta)} +
\frac{\beta}{\sin^2 (k \theta)}\right]\,,
\label{potTTWbycorrect}
\end{eqnarray}
where $k=m/n$ and $m$ and $n$ are two integers (with no common divisors)\cite{TTW:2009, TTW:2010}. The other
is the PW potential\cite{PostWinternitz:2010}
\begin{eqnarray}
V_{\rm PW} = \frac{a}{r} + \frac{1}{r^2}\left[\frac{\mu}{\cos^2(\frac{k}{2} \theta)} + \frac{\nu}{\sin^2(\frac{k}{2} \theta)} \right]\,,
\label{potPWbycorrect}
\end{eqnarray}
where again $k=m/n$ is rational.

In the TTW case (\ref{potTTWbycorrect}) the lowest order of an additional
integral $Z$ was shown to be\cite{QuesneTTW:2010,Pogosyan2011, Gonera, KalninsKM:2010}
\begin{eqnarray}
N = 2 (m + n - 1)\,,
\label{intorderbycorrect}
\end{eqnarray}
both in classical and quantum mechanics.

The TTW and PW potentials are related by coupling constant
metamorphosis\cite{PostWinternitz:2010, HietarintaPRL:1984, Boyer, 2010JPhA...43c5202K}.
This transformation takes integrable and superintegrable systems into other systems 
that are also integrable or superintegrable, respectively.
However, in general the order of the integrals of motion is not preserved,
nor is their polynomial character. Special conditions\cite{2010JPhA...43c5202K} must
be satisfied for polynomial integrals to be transformed into polynomials of the
same order. It has been conjectured and verified for $k = 1$ and $2$ that these
conditions are satisfied for the TTW and PW systems\cite{PostWinternitz:2010}.

Thus we may expect that in any study of superintegrable systems separating in
polar coordinates in $E_2$ with an additional integral of order $N$ with
$2\leq N < \infty$ the TTW and PW systems will appear as ``standard'' potentials.

A further comment is in order here and should be taken into account in any
classification of superintegrable systems. If a certain potential $V(x, y)$ appears
as superintegrable in $E_2$ with two integrals $X_1$ and $X_2$ that are polynomials
of order $m$ and $n$, respectively, then the same potential will reappear at higher orders.
Indeed $X_1 X_2 + X_2 X_1$ will be an integral of order $m + n$ for the
same Hamiltonian. The commutator $[X_1, X_2]$ (or Poisson bracket $\{X_1, X_2\}_{\rm P. B.}$)
will be a polynomial integral of order $m + n - 1$. Such ``trivially'' superintegrable systems should
be weeded out of lists of new superintegrable systems.

For fourth order superintegrable potentials, in addition to the Hamiltonian $H$, we have two more conserved quantities.
A second-order integral (\ref{X}), associated with the separation of
variables in polar coordinates for which the potential takes the form
\begin{equation}
V(r,\,\theta) \ = \ R(r) \ + \ \frac{S(\theta)}{r^2} \ ,
\end{equation}
and a fourth-order one
\begin{align}
\label{Y}
Y &\ =\  A_1\,\{L_z^3,\ p_x \} \ + \ A_2\,\{L_z^3,\ p_y \} \ + \ A_3\,\{L_z,\
p_x\,(p_x^2+p_y^2) \} \ + \ A_4\,\{L_z,\ p_y\,(p_x^2+p_y^2)    \} \\ &
+ \ B_1\,(p_x^4-p_y^4) \ + \ 2\,B_2\,p_x\,p_y\,(p_x^2+p_y^2) \ + \ B_3\,
\{L_z^2,\  p_x^2-p_y^2  \} \ + \ 2\,B_4\,\{L_z^2,\ p_x\,p_y \}
\nonumber \\ &
+ \ C_1\,\{L_z,\ 3\,p_x^2\,p_y-p_y^3 \} \ + \ C_2\,\{L_z,\ p_x^3 -3\,p_y^2\,p_x \} \
+ \ D_1\,(p_x^4+p_y^4-6\,p_x^2\,p_y^2)
\nonumber \\ &
 + \ 4\,D_2\,p_x\,p_y\,(p_x^2-p_y^2)
\ + \ \{ g_1(x,y),\,p_x^2  \} \ + \  \{ g_2(x,y),\,p_x\,p_y  \}
\ + \  \{ g_3(x,y),\,p_y^2  \}
\ + \  g_4(x,y)\ ,\nonumber
 \end{align}
where the $A_i,\,B_i,\,C_i$, $D_i$ are real constants. The bracket $\{\cdot ,\, \cdot   \}$ denotes an anticommutator and $R(r),\, S(\theta),\, g_{1,2,3,4}(x,y)$ are real functions such that
\begin{equation}
[H,\,Y] \ = \ [H,\,X] \ = \ 0   \ .
\end{equation}
Under rotations around the $z$-axis, each of the six pairs of parameters in (\ref{Y})
\[
(A_1,A_2),\ (A_3,A_4),\ (B_1,B_2),\ (B_3,B_4),\ (C_1,C_2),\ (D_1,D_2)\ ,
\]
forms a doublet. In general, $Y$ may contain the $O(2)$ singlets but by linear combinations of the form $Y
+ u_1\, X^2 + u_2\, H^2 + u_3\, X  H$, where the $u_i$ are constants, we
eliminated all such trivial terms. Under rotations through the angle $\theta$ the doublets $A_i,\,B_i,\,C_i$, $D_i$
rotate through $\theta,\,2\theta,\,3\theta$, $4\theta$, respectively.

We have $[Y,X]=C \neq 0$ where $C$ is in general a $5^{th}$-order linear operator.
We thus obtain a finitely generated polynomial algebra of integrals
of motion. We are looking for fourth-order superintegrable systems, so at least one of the constants
$A_i,\,B_i,\,C_i,\,D_i$ is different from zero. The operator $Y$ is the most general polynomial expression for a fourth-order Hermitian operator of the required form. The commutator $[H,\,Y]$ contains derivatives of order up to three.

The operator $Y$ in (\ref{Y}) is given in Cartesian coordinates for brevity. Putting
\[
p_x \ = \ -i\,\hbar\,(\cos \theta \,\partial_r - \frac{\sin \theta}{r}\,\partial_\theta) \ , \qquad p_y \ = \ -i\,\hbar\,(\sin \theta \,\partial_r + \frac{\cos \theta}{r}\,\partial_\theta) \ ,
\]
we obtain the corresponding expression in polar coordinates.
Explicitly, the leading terms of the integral $Y$ are
\begin{equation}
\begin{aligned}
\label{Ylead}
Y &\ = \ \hbar^4\,\bigg( ( B_1 \cos 2 \theta + B_2 \sin 2 \theta + D_1 \cos 4 \theta   + D_2
\sin 4 \theta)\,\partial^4_r
+  \frac{1}{ r^4}  \bigg[   D_2 \sin 4 \theta
\\ & + D_1 \cos 4 \theta  -2 r \left(A_{1} r^2+A_4\right) \sin \theta
  - \left(B_2+2\,B_{4} r^2\right) \sin 2 \theta +2 r \left(A_2 r^2+A_4\right)
\cos \theta
\\ &
 - \left(B_1+2\,B_3 r^2\right) \cos 2 \theta  -2\,r\, (
  \,C_1\,  \cos 3 \theta-\,C_2\,  \sin 3 \theta)    \bigg]\,  \partial^4_\theta \\ &
- \frac{2}{r^2} \bigg[
3(
  D_{1} \cos 4\theta
+ D_{2} \sin 4\theta
)
- r^2 (
  B_{3}   \cos 2\theta
+ B_{4}  \sin 2\theta )
\\ &
+ r   (
A_{3} \sin \theta - A_4 \cos  \theta
-3(\,C_1 \, \cos 3\theta
- \,C_2 \sin 3\theta)\,  )
\bigg] \partial^2_r\,\partial^2_\theta
\\ &  - \frac{2}{ r} \bigg[   B_1 \sin 2 \theta- B_2 \cos 2 \theta +2 (D_1 \sin
4 \theta  - D_2 \cos 4 \theta)
 -r\,(\,C_1\, \sin 3 \theta + \,C_2\, \cos 3 \theta
\\ &  + A_3 \cos \theta + A_4 \sin \theta      )
\bigg]\,\partial^3_r\,\partial_\theta
-\frac{2}{r^3}  \bigg[    B_1 \sin 2 \theta- B_2 \cos 2 \theta
\\ &   -2 \,(D_1 \sin 4 \theta + D_2 \cos 4 \theta)
  - r (A_3 \cos \theta+ \,A_4 \sin \theta   -3 ( \,C_1\,
\sin 3 \theta + C_2\, \cos 3 \theta ) \ )
\\ &     + 2\,r^2 \left( B_3 \sin 2 \theta - B_{4} \cos 2 \theta \right) -\, r^3
\left( A_{1} \cos \theta +A_2 \sin \theta \right)
  \bigg]\, \partial_r\,\partial^3_\theta \ \bigg)  + \text{lower order terms}  \ .
\end{aligned}
\end{equation}
For convenience let us introduce the functions
\[
G_1(r,\,\theta) \ = \ g_1\,\cos^2\theta +
g_2\,\sin^2\theta+g_3\,\cos\theta\sin\theta \ ,
\]
\[
G_2(r,\,\theta) \ = \ \frac{
g_1\,\sin^2\theta+ g_2\,\cos^2\theta-g_3\,\cos\theta\sin\theta}{r^2} \ ,
\]
\[
G_3(r,\,\theta) \ = \ -\frac{
g_1\,\sin2\theta-g_2\,\sin2\theta-g_3\,\cos2\theta}{r} \ ,
\]
\begin{equation}
\label{Gfunctionsg}
G_4(r,\,\theta) \ = \ g_4 \ .
\end{equation}
Then, we can write
\begin{equation}
\begin{split}
& \{ g_1(x,y),\,p_x^2  \} \ + \  \{ g_2(x,y),\,p_x\,p_y  \}
\ + \  \{ g_3(x,y),\,p_y^2  \}
\ + \  g_4(x,y)\
\\&
=   \ -\hbar^2\,(\{ G_1(r,\theta),\,\partial_r^2  \} \ + \  \{ G_3(r,\theta),\,\partial_r\,\partial_\theta  \}
\ + \  \{ G_2(r,\theta),\,\partial_\theta^2  \})
\ + \  G_4(r,\theta)\,.
\end{split}
\end{equation}

\section{DETERMINING EQUATIONS FOR A FOURTH-ORDER INTEGRAL}
\label{DETERMINING EQUATIONS}

\subsection{Commutator $[H,\,Y]$}
Since $Y$ is a fourth-order operator the commutator $[H,\,Y]$ would be a fifth-order one, i.e. we have
\begin{equation}\label{}
  [H,\,Y] \ = \ \sum^{N+1}_{k+l=0} Z_{k,l}(r,\,\theta)\,\frac{\partial^{k+l}}{\partial r^k\,\partial \theta^l} \ , \qquad N=4
\end{equation}
and we require $Z_{k,l}=0$ for all $k$ and $l$. The terms of order $k+l=5$ and $k+l=4$ vanish automatically. It was already shown that for arbitrary $N$ the terms of order $N+1$ and $N$ vanish \cite{PostWinternitz:2015}. Moreover, only terms with $k+l$ of the opposite parity than $N$ provide independent determining equations ($Z_{k,l}=0$). Those with the same parity provide equations that are differential consequences of the first ones.

Finally, we obtain
{
\begin{align}
[H,Y]   &=
{\cal A}_{rrr}\,\frac{\partial^3  }{ \partial r^3}\  + \
{\cal A}_{rr\theta}\,\frac{\partial^3  }{\partial r^2\partial\theta}\ + \
{\cal A}_{r\theta\theta}\,\frac{\partial^3  }{\partial r\partial \theta^2}\ + \
{\cal A}_{\theta\theta\theta}\,\frac{\partial^3  }{ \partial \theta^3}
\nonumber \\&
 +{\cal C}_{r}\,\frac{\partial  }{\partial r}\ + \
 {\cal C}_{\theta}\,\frac{\partial  }{\partial\theta} \   = 0\  ,
\label{commutator}
\end{align}
}
here the coefficients ${\cal A}_{rrr}, {\cal A}_{rr\theta},...$, are real
functions of $r$ and $\theta$. The determining equations are obtained by setting all coefficients in (\ref{commutator}) equal to zero. We thus obtain a total of 6 PDEs for the functions $G_1,\,G_2,\,G_3,\,G_4$, $R$ and $S$. Four equations are independent of $\hbar$
\begin{align}
 G_{1}^{(1,0)} & \, = \,     F_1\, R'  -\frac{2 \, F_1}{r^3}\,S   +  \frac{F_2
}{r^2}S'
\  ,
\label{Eq30p} \\
 \frac{1}{r^2}
 \Big(
G_{2}^{(0,1)}  + \frac{1}{r} G_{3}
  \Big) & \, = \,    F_3\, R'  -\frac{2 \, F_3}{r^3}\,S   + \frac{F_4 }{r^2}S'
\ ,
\label{Eq03p} \\
 \frac{1}{r^2}G_{1}^{(0,1)} +  G_{3}^{(1,0)}
& \, =  \,    3\,F_2\, R'  -\frac{6 \, F_2}{r^3}\,S      +\frac{F_5}{r^2}S'   \
,
\label{Eq21p} \\
  \frac{2}{r^3} G_1
+  G_{2}^{(1,0)}
+ \frac{1}{r^2} G_{3}^{(0,1)}
& \,  =  \,   F_5\, R'   -\frac{2 \, F_5}{r^3}\,S    +\frac{3\,F_3 }{r^2}S' \ .
 \label{Eq12p}
\end{align}
They are the same for classical or quantum systems, and two equations contain quantum corrections proportional to $\hbar^2$
\begin{equation}    
\begin{aligned}
\label{Eq10p}
&G_3 \,\frac{1}{r^2}\,S'
+ 2 \,G_1\, ( R'- \frac{2}{r^3}\,S) -\frac{1}{2} G_4{}^{(1,0)}
\\ &
+ \hbar^2\bigg[\frac{G_1{}^{(0,2)}}{2 r^3}-\frac{3 G_3{}^{(0,1)}}{4 r^2}
+\frac{G_3{}^{(0,3)}}{4 r^2}-\frac{G_1{}^{(1,0)}}{r^2}-\frac{3}{2} G_2{}^{(1,0)}
+\frac{5 G_3{}^{(1,1)}}{4 r}
\\ &
+\frac{G_1{}^{(1,2)}}{2 r^2}
+\frac{1}{2} G_2{}^{(1,2)}
+\frac{2 G_1{}^{(2,0)}}{r}
-\frac{1}{2} r G_2{}^{(2,0)}
+\frac{3}{4} G_3{}^{(2,1)}
+G_1{}^{(3,0)}\bigg] \\ &\hspace{30pt}
= \ \hbar^2\bigg[
 \left(\frac{6 F_6}{r^4}-\frac{24 F_1}{r^5}-\frac{2 F_{11}}{r^3}\right)\,S
 +\left(\frac{6 F_{10}}{r^4}-\frac{4 \,F_7}{r^3}+\frac{F_{12}}{r^2}\right) S'
\\ &  \hspace{30pt}
+\left(\frac{F_8}{r^2}-\frac{2 \,F_5}{r^3}\right)\, S''
+\frac{F_3}{r^2}\,S'''
+ F_{11}\, R'
+F_6\, R''
+ F_1 \,R'''
 \bigg] \ ,
\end{aligned}
\end{equation}

\vspace{0.4cm}

\begin{equation}  
\begin{aligned}
\label{Eq01p}
& G_3 \,( R'- \frac{2}{r^3}\,S) \  +  \ 2 \,G_2\,\frac{1}{r^2}\,S' \ - \ \frac{G_4{}^{(0,1)}}{2 r^2}
\\ &
+\hbar^2\bigg[
 \frac{G_3}{4 r^3}
-\frac{G_2{}^{(0,1)}}{r^2}
+\frac{5 G_3{}^{(0,2)}}{4 r^3}
+\frac{G_2{}^{(0,3)}}{r^2}
-\frac{G_3{}^{(1,0)}}{4 r^2}
+\frac{G_1{}^{(1,1)}}{r^3}      \\ &
+\frac{3 G_3{}^{(1,2)}}{4 r^2}
+\frac{G_3{}^{(2,0)}}{2 r}
+\frac{G_1{}^{(2,1)}}{2 r^2}
+\frac{1}{2} G_2{}^{(2,1)}
+\frac{1}{4} G_3{}^{(3,0)}
\bigg]
\hspace{30pt}
\\ &  = \  \hbar^2\bigg[
\left(\frac{6 \,F_7}{r^4}-\frac{24 \,F_2}{r^5}-\frac{2\, F_{12}}{r^3}\right)\,S
+\left(\frac{6\, F_5}{r^4}-\frac{4\, F_8}{r^3}+\frac{F_{13}}{r^2}\right)\, S'
\\ &  \hspace{30pt}
+\left(\frac{F_9}{r^2}-\frac{6 F_3}{r^3}\right) \,S''
+ \frac{F_4}{r^2}\, S'''
+ F_{12}\, R'
+  F_7 \,R''
+  F_2\,R'''
\bigg]
\ .
\end{aligned}
\end{equation}

\vspace{0.25cm}
The functions $F_1 \ldots F_{13}$ are completely determined by the constants
$A_i,B_i,C_i,D_i$ figuring in the leading part of the integral $Y$. They are presented in the Appendix.

The system of equations (\ref{Eq30p})-(\ref{Eq01p}) is the main object of study of the present paper.

\clearpage

\subsection{Linear compatibility condition}
The compatibility equation for (\ref{Eq30p})-(\ref{Eq12p}) is a linear equation 
{\small
\begin{equation}
\begin{aligned}
& 0 \ = \frac{1}{r^2}
  \Big(
16\, \Big[
A_3 \,\cos \theta 
+ A_4 \,\sin \theta 
- 9\,( C_1\, \sin  3\,\theta
+ C_2\, \cos  3\,\theta)
\Big] \, S
\\&
- 20\, \Big[
A_{{4}} \,\cos \theta
- A_{{3}} \,\sin \theta
+ 9\,(C_2\,\sin  3\,\theta
- C_1\,\cos  3\,\theta)
\Big] \, S'
\\&
+ 80\, \Big[
\,C_1 \,\sin 3\,\theta
+\,C_2 \,\cos 3\,\theta
\Big] \,S''
\\ &
+ 5\, \Big[
A_{{3}}\,\sin  \theta
-A_{{4}}\,\cos  \theta
+ 3(\,C_2 \,\sin 3\,\theta
- C_1 \,\cos 3\,\theta)
\Big] \,  S^{(3)}
 \\ &
 - \,\Big[
A_{{3}} \,\cos \theta 
+ A_{{4}} \,\sin \theta
+ C_1 \,\sin  3\,\theta
+ C_2 \,\cos  3\,\theta
\Big] \,   S^{(4)}
\Big)
\\ &
+ \frac{1}{r^3}
  \Big(
96\, \Big[
B_2 \,\cos 2\,\theta 
- B_1 \,\sin 2\,\theta 
+ 8\,( D_1\, \sin  4\,\theta
- D_2\, \cos  4\,\theta)
\Big] \, S
\\&
+ 40\, \Big[
B_{{1}} \,\cos 2\,\theta
+ B_{{2}} \,\sin 2\,\theta
- 20\,(D_2\,\sin  4\,\theta
+ D_1\,\cos  4\,\theta)
\Big] \, S'
\\&
+ 20\, \Big[
\,(B_2 \,\cos 2\,\theta
-\, B_1 \,\sin 2\,\theta) + 
14\, (D_2\,\cos  4\,\theta
- D_1\,\sin 4\,\theta)
\Big] \,S''
\\ &
+ 10\, \Big[
(B_{{2}}\,\sin 2\,\theta
+ B_{{1}}\,\cos 2\,\theta )
+ 4\, (D_1 \,\cos 4\,\theta
+ D_2 \,\sin 4\,\theta)
\Big] \,  S^{(3)}
 \\ &
 - \,\Big[
B_{{2}} \,\cos 2\,\theta 
- B_{{1}} \,\sin 2\,\theta
+ 2\, (D_2 \,\cos 4\,\theta
- D_1 \,\sin 4\,\theta
\Big] \,   S^{(4)}
\Big)
\\&
+ \Big[
24\, r^3 \, \big(A_{{1}}\, \cos  \theta
 + A_{{2}} \, \sin  \theta \big)
- 12\, {r}^2 \big(
B_{{4}} \cos  2\,\theta - 
B_{{3}}  \sin  2\,\theta \big)
 \\ &
- r\, \big(2\, (A_{{4}} \,\sin  \theta
+ A_{{3}} \,\cos  \theta )
+ 18\, (C_1 \,\sin  3\,\theta
+ C_2 \,\cos  3\,\theta) \big)
\\ &
+ 3\, (B_{{2}} \cos  2\,\theta
-  B_{{1}} \sin  2\,\theta)
+ 30\, (D_{{2}} \cos  4\,\theta
-  D_{{1}} \sin  4\,\theta)
 \Big] \, R'
  \\ &
+
 \Big[
 36 \,{r}^{4}\big(
A_{{1}} \,\cos  \theta 
+  A_{{2}} \,\sin  \theta \big)
 - 12 \,{r}^3\, \big(
  B_{{3}}  \sin  2\,\theta
-  B_{{4}}  \cos  2\,\theta\big)
 \\ &
 + r^2\,\big(2\, 
(A_{{3}}\,\cos \theta + A_{{4}}\,\sin \theta)  
+ 18 \,( C_1  \,\sin 3\,\theta
+  C_2  \,\cos 3\,\theta)
 \big) \\&
+ r\,\big(3\, (B_{{1}} \,\sin  2\,\theta
- B_{{2}} \,\cos  2\,\theta))
- 30\, (D_{{2}} \,\cos  4\,\theta
- D_{{1}} \,\sin  4\,\theta) \big)
 \Big] \, R''
 \\&
+
 \Big[12\, {r}^{5}
\big(
A_{{1}} \,\cos  \theta 
+  A_{{2}} \,\sin  \theta
\big)
    + 14\,{r}^4  \big(
      B_{{4}}\cos  2\,\theta
    - B_{{3}}\sin  2\,\theta  \big)
 \\&
  + r^3\, \big(5\,(A_{{3}}\,\cos  \theta
  + A_{{4}} \sin \theta)
- 3\, (C_1 \,\sin 3\,\theta
+  C_2 \,\cos 3\,\theta)\big)
 \\&
 + 12\,r^{2}\,(D_{{2}}\,\cos  4\,\theta
 - D_{{1}}\,\sin  4\,\theta) \
 \Big]  R^{(3)}
 \\ &
+ 
 \Big[
    {r}^{6}\big(
A_{{1}} \,\cos  \theta
   +  A_{{2}} \,\sin  \theta
\big)
    + 2\,{r^5}\,  \big(
      \, B_{{4}}\cos  2\,\theta
    - \, B_{{3}}\sin  2\,\theta  \big)
\\&
  + r^4\,\big(
  A_{{4}} \sin \theta
+ A_{{3}}\,\cos  \theta
- 3\, (C_1 \,\sin 3\,\theta
+ C_2 \,\cos 3\,\theta)
 \big)
\\&
 - r^3\, \big( B_{{1}}\,\sin  2\,\theta
 - B_{{2}}\,\cos  2\,\theta
 + 2\,(D_{{2}}\,\cos  4\,\theta
 - D_{{1}}\,\sin  4\,\theta) \big) \
 \Big]  R^{(4)} \ ,
\label{CCsep}
\end{aligned}
\end{equation}}
$\!\!$for the functions $R(r)$ and $S(\theta)$ in the potential. It is the same
in classical and quantum mechanics (as are equations (\ref{Eq30p}), ..., (\ref{Eq12p})).

We will henceforth define \emph{exotic potentials} as those
for which the compatibility condition is satisfied trivially.
For $N = 4$ this means all the coefficients in the integral $Y$
of (\ref{Y}) except $A_1, A_2, B_3$ and $B_4$ vanish. \emph{Standard
potentials} must satisfy the linear PDE (\ref{CCsep}) nontrivially.
Hence at least one of the constants $A_i, B_i, C_i$ or $D_i$ in (\ref{CCsep})
must be nonzero.


Differentiating (\ref{CCsep}) with respect to $r$ we eliminate $S(\theta)$ for the equation. Following paper\cite{ELW} we obtain $8$ ODEs for $R(r)$ by setting the coefficients of $\cos k\theta$ and $\sin k\theta$ for $k=1,2,3,4$ equal to zero. These together with the determining equations imply that the only possible forms of $R(r)$ are
\begin{equation}\label{Rp}
  R(r) \ =\ 0 \ , \qquad R(r) \ =\ \frac{a}{r} \ , \qquad R(r) \ =\ b\,r^2 \ .
\end{equation}

The result (\ref{Rp}) is actually a consequence of Bertrand's theorem\cite{Bertrand} and is valid
for integrals of motion $Y$ of any order\cite{EWY}. Let us now consider each of the cases
listed in (\ref{Rp}) separately.

\section{NON CONFINING POTENTIAL    $V(r,\theta)=\frac{S(\theta)}{r^2}$
}
\label{Potentials}
Here we address the case of a non confining potential, namely $R(r)=0$. It turns out that this case provides the most general form of the function $S(\theta)$. The same function $S(\theta)$ will reoccur for $R(r)\neq 0$.

The equations (\ref{Eq30p}), (\ref{Eq21p}) and (\ref{Eq12p}) corresponding
to the determining equations ${\cal A}_{rrr}={\cal A}_{rr\theta}={\cal
A}_{r\theta\theta}=0$, respectively,  take the following
form
\begin{equation}
\begin{split}
 &
 G_{{1}}^{(1,0)} \left( r,\theta \right) {r}^{3}  -
\bigg(
\Big(
  A_{{3}} \cos   \theta
+ A_{{4}} \sin   \theta
+ C_{{1}}\sin   3\,\theta
+ C_{{2}}\cos   3\,\theta
\Big) r
\\& -
\Big(
     B_{{1}}\sin   2\,\theta
-    B_{{2}}\cos   2\,\theta
+ 2\,\left(
     D_{{1}}\sin   4\,\theta
-    D_{{2}}\cos   4\,\theta
\right)
\Big)
\bigg)\, S'
\\& + 4 \bigg(
 \,  B_{{1}} \cos   2\,\theta
+\,  B_{{2}} \sin   2\,\theta
+\,  D_{{1}} \cos   4\,\theta
+\,  D_{{2}} \sin   4\,\theta
 \bigg)\,S  \ = \ 0\ ,
\label{G10}
\end{split}
\end{equation}
\begin{equation}
 \begin{split}
&
G_{{3}}^{(1,0)} \left( r,\theta \right)  {r}^{4}
+
G_{{1}}^{(0,1)} \left( r,\theta\right)   {r}^{2}
-2
\bigg(
  \Big(
  \,B_{{3}} \cos   2\,\theta
+ \,B_{{4}} \sin   2\,\theta
\Big)    {r}^{2}
\\&
- \Big(
  \,A_{{3}}\sin   \theta
- \,A_{{4}}\cos   \theta
-3\,\left(
 \,C_{{1}}\cos   3\,\theta
-\,C_{{2}}\sin   3\,\theta
\right)
\Big)    r
\\&
  - 3 \left(
 \,D_{{1}}\cos   4\,\theta
+\,D_{{2}}\sin   4\,\theta
\right)
\bigg) \,S'
\\&
+
6\bigg(
 \left(
  A_{{3}} \,\cos   \theta
+ A_{{4}} \,\sin   \theta
+ C_{{1}} \,\sin   3\,\theta
+ C_{{2}} \,\cos   3\,\theta
\right)\,   r
\\&
- \left(
   \,B_{{1}}\sin   2\,\theta
-  \,B_{{2}}\cos   2\,\theta
+ 2 \left(\,D_{{1}}\sin   4\,\theta
- \,D_{{2}}\cos   4\,\theta    \right)
\right)
\bigg)\,S \
=  \  0\ ,
\label{G30}
 \end{split}
\end{equation}
\begin{equation}
 \begin{split}
&
G_{{2}}^{(1,0)} \left( r,\theta \right) {r}^{5}
+
G_{{3}}^{(0,1)} \left( r,\theta \right) {r}^{3}
+2\,G_{{1}} \left( r,\theta \right) {r}^{2}
\\&
-3\Bigg(
 \left(
 \,A_{{1}}\cos   \theta
+\,A_{{2}}\sin   \theta
\right) {r}^{3}
+ 2\left(
 \,B_{{4}}\cos   2\,\theta - \,B_{{3}}\sin   2\,\theta
\right) {r}^{2}
\\&
+ \left(
A_{{3}} \cos  \theta
+\,A_{{4}}\sin   \theta
-3\,(C_{{1}}\sin   3\,\theta    + \,C_{{2}}\cos   3\,\theta  ) \right) r
\\&
 -\,B_{{1}}\sin   2\,\theta
 +\,B_{{2}}\cos   2\,\theta
+2\,\left(
   D_{{1}}\sin   4\,\theta
-\,D_{{2}}\cos   4\,\theta
\right)
\Bigg)\,S'
\\&
+
4\Bigg(
  \left(
  \,B_{{3}} \cos   2\,\theta
+ \,B_{{4}} \sin   2\,\theta
\right) {r}^{2}
\\&
+ \left(
 \,A_{{4}} \cos   \theta
- \,A_{{3}} \sin   \theta
+ 3\,\left(
  C_{{1}}\cos  3\,\theta
- C_{{2}}\sin  3\,\theta
\right)
\right) r
\\&
-3\,\left(
 \,D_{{1}}\cos   4\,\theta
+\,D_{{2}}\sin   4\,\theta
\right)
\Bigg)\, S  \  = \  0\ .
\label{G20}
 \end{split}
\end{equation}
The above equations can be integrated to find the functions
$G_{1}(r,\theta)$,
$G_{3}(r,\theta)$ and  $G_{2}(r,\theta)$ in terms of $S(\theta)$. They are given by the following
formulas:
\begin{equation}
\begin{split}
  G_{{1}} \left( r,\theta \right) & =
\Bigg(
{\frac {1}{2\,{r}^{2}}}  \Big(
{B_{{1}}\sin\,2\theta -  B_{{2}}\cos\,2\theta}
+  2 \left(
{D_{{1}}\sin\,4\theta -  D_{{2}}\cos\,4\theta}
\right)\,\Big)
\\ &
-{\frac {1}{r}}\Big( {A_{{3}}\cos\,\theta +  A_{{4}}\sin\,\theta}
+ {C_{{1}}\sin\,3\theta +  C_{{2}}\cos\,3\theta}  \Big)
\Bigg)\, S'
\\&
+ \,{\frac {2}{{r}^{2}}}  \Big(
{B_{{1}}\cos\,2\theta\, +  B_{{2}}\sin\,2\theta\,}
 +
 {D_{{1}}\cos\,4\theta\, +  D_{{2}}\sin\,4\theta\,}
  \Big)\,S
+ \gamma_{{1}} \left(\theta \right)\ ,
\label{G1gen}
\end{split}
\end{equation}
\begin{equation}
 \begin{split}
 G_{{3}} \left( r,\theta \right) &=
 -{\frac {2}{r}}
 \left(  {B_{{3}}\cos\,2\theta\, +  B_{{4}}\sin\,2\theta\,} \right)  \,S'
 +{\frac {1}{{r}^{2}}}\Big[
 3\,\big(  \doubletA + \doubletC  \big) S
\\&
+ \frac{3}{2} \Big(
({A_{{3}}\sin\,\theta\, -  A_{{4}}\cos\,\theta\,})
 - 3 (C_{{1}}\cos\,3\theta\, -  C_{{2}}\sin\,3\theta\,)
 \Big)  S'
\\&
-\frac{1}{2}\, \Big(
( {A_{{3}}\cos\,\theta\, +  A_{{4}}\sin\,\theta\,} )  + \,
( {C_{{1}}\sin\,3\theta\, +  C_{{2}}\cos\,3\theta\,} )  \,
\Big) S''
 \Big]
\\&
+\frac {1}{{r}^{3}} \left[
-\frac{10}{3} \Big(
( {B_{{1}}\sin\,2\theta\, -  B_{{2}}\cos\,2\theta\,} ) +
2({D_{{1}}\sin\,4\theta\, -  D_{{2}}\cos\,4\theta\,} )
 \Big) S
\right.\\&\left.
 + \Big(
 {B_{{1}}\cos\,2\theta\, +  B_{{2}}\sin\,2\theta\,}
  + 4\,({D_{{1}}\cos\,4\theta\, +  D_{{2}}\sin\,4\theta\,}
 )\Big) S'
 \right.\\&\left.
 + \frac{1}{6} \Big(
 {B_{{1}}\sin\,2\theta\, -  B_{{2}}\cos\,2\theta\,}
 +  2 ( {D_{{1}}\sin\,4\theta\, -  D_{{2}}\cos\,4\theta\,})
 \Big) S''
 \right]
\\&
+ \frac{1}{r}{ \gamma_{{1}} ' \left( \theta \right) }
+ \gamma_{{3}} \left( \theta\right)
 \ ,
\label{G3gen}
 \end{split}
\end{equation}

\begin{equation}
 \begin{split}
 G_{{2}} \left( r,\theta \right) & =
 - {\frac {3}{r}} \left(
    \,A_{{1}} \cos \theta   +\,A_{{2}}\sin \theta
  \right)  S'
+{\frac {1}{{r}^{2}}}  \Big[
\left( 2\,(\doubletBB)   \right) S
\\&
 + 5\left(
  \,B_{{3}}\sin   2\,\theta
 -\,B_{{4}}\cos  2\,\theta
 \right) S'
- \left(
 B_{{3}} \cos 2\,\theta
+B_{{4}}\sin 2\,\theta
 \right) S''
 \Big]
\\&
+
\frac {1}{ {r}^{3}} \Big[
-\frac{7}{3}\Big(
\left(
 \,A_{{3}}\sin   \theta
-\,A_{{4}} \cos  \theta
\right)
-3\left(
 \, C_{{1}}\cos   3\,\theta
-\, C_{{2}}\sin   3\,\theta
\right)
\Big) S
\\&
 -\frac{1}{6} \Big(
 \left( \,A_{{3}}\cos   \theta  +\,A_{{4}}\sin   \theta  \right)
       -{47}  \left(\,C_{{1}}\sin  3\,\theta  +  \,C_{{2}}\cos  3\,\theta  \right)
  \Big) S'
  \\&
  +  \frac{2}{3}
  \Big( \left( \,A_{{3}} \sin  \theta   - \,A_{{4}} \cos   \theta
  \right)
  -  3\left(   \,C_{{1}}\cos    3\,\theta   - \,C_{{2}} \sin   3\,\theta    \right)
  \Big) S''
  \\&
  -\frac{1}{6} \Big(
 \left(   \,A_{{3}} \cos   \theta  + \,A_{{4}}\sin   \theta  \right)
 +
 \left(  \,C_{{1}}\sin   3\,\theta  +\,C_{{2}}\cos  3\,\theta  \right)
  \Big) S'''
\Big]
\\&
+ \frac {1}{{r}^{4}} \Big[
-\frac{2}{3} \Big(
\left( \,B_{{1}}\cos  2\,\theta  +\,B_{{2}}\sin  2\,\theta   \right)
 + 13 \left( \,D_{{1}}\cos   4\,\theta   +\,D_{{2}}\sin   4\,\theta    \right)
 \Big) S
 \\&
 -\frac{1}{3} \Big(
 \left(  \,B_{{1}}\sin  2\,\theta   -\,B_{{2}}\cos 2\,\theta  \right)
  + {20} \left( \,D_{{1}}\sin  4\,\theta  - \,D_{{2}}\cos   4\,\theta  \right)
 \Big) S'
 \\&
 + \frac{1}{3} \Big(
    \left(\,B_{{1}}\cos 2\,\theta  +\,B_{{2}}\sin  2\,\theta \right) +
 {4}\left(  \,D_{{1}}\cos   4\,\theta   + \,D_{{2}}\sin   4\,\theta \right)
 \Big)  S''
\\&
+ \frac{1}{24}\Big(
\left( \,B_{{1}}\sin \  2\,\theta  - \,B_{{2}}\cos  2\,\theta \right)
+
{2}\left( \,D_{{1}}\sin   4\,\theta  - \,D_{{2}}\cos  4\,\theta \right)
\Big) S'''
 \Big]
 \\&
 + \frac{1}{r} \gamma_{{3}}' (\theta)
 + \frac{1}{r^2} \gamma_{{1}} (\theta)
 + \frac{1}{2 r^2} \gamma_{{1}}'' (\theta)
 + \gamma_{{2}} \left( \theta\right)  \ ,
\label{G2gen}
 \end{split}
\end{equation}
where $\gamma_{{1}} \left( \theta \right), \gamma_{{2}} \left( \theta \right)$
and $\gamma_{{3}} \left( \theta \right)$ are functions of $\theta$ still to be determined.

The determining equation ${\cal A}_{\theta\theta\theta}=0$ (\ref{Eq03p}) has not been used yet. We now substitute the expressions (\ref{G1gen})-(\ref{G2gen}) into (\ref{Eq03p}) in order to obtain further information on functions of $\theta$, namely $S(\theta)$ and $\gamma_{k}(\theta)$, \,$k=1,2,3$. The variable $r$ still figures in the obtained result but only explicitly as $r^{k}$, $k=0,1,2,3,4$ (no unknown functions of $r$ remain). Thus, the coefficient of $r^k$ in ${\cal A}_{\theta\theta\theta}=0$ (\ref{Eq03p}) must vanish for each $k$ and we obtain a system of $5$ equations for the $4$ remaining functions of $\theta$. They are
\begin{equation}
 \begin{split}
 \label{gamma2}
 0\ = \   \gamma'_{{2}}  \ ,
\end{split}
\end{equation}
 \begin{equation}
 \begin{split}
 \label{eqa1}
 0\ =& \
 4\,\Big( \,A_{{1}}\cos \,\theta   + \,A_{{2}} \sin  \theta
 \Big) S
 + 14 \Big( \,A_{{1}}\sin \,\theta   - \,A_{{2}} \cos \theta
 \Big) S'
 \\&
  -6 \Big( \,A_{{1}}\cos  \theta  + \,A_{{2}} \sin \theta \Big) S''
  + 2\, \gamma''_{{3}}   +2\,\gamma_{{3}} \ ,
   \end{split}
\end{equation}

 \begin{equation}
 \begin{split}
 \label{eqb3}
 0=&
-16 \Big(
  \,B_{{3}} \sin 2\,\theta
 -\, B_{{4}}\cos  2\,\theta
 \Big) S   +
  28 \Big(\,
B_{{3}} \cos 2\,\theta + \,B_{{4}} \sin  2\,\theta
 \Big) S'
 \\&
 + 14 \Big(
   \,B_{{3}} \sin  2\,\theta   -\,B_{{4}}\cos  2\,\theta
 \Big) S''
 -2  \Big( \,B_{{3}} \cos   2\,\theta  + \,B_{{4}}\sin   2\,\theta
 \Big) S^{(3)}
 \\&
 +4\, \gamma'_{{1}}   +    \gamma^{(3)}_{{1}} \ ,
 \end{split}
\end{equation}

 \begin{eqnarray}
 0 \ &=& \ 
96\,\Big(
\,(B_{{2}}\cos \,2\,\theta - \,B_{{1}}\sin \,2\,\theta )
+ 8\, (D_{{1}}\sin \,4\,\theta -\,D_{{2}}\cos \,4\,\theta)
\Big)
S
\nonumber \\
&+& 40\,\Big(
 \,(B_{{1}}\cos \,2\,\theta + \,B_{{2}}\sin \,2\,\theta)
- 20 (\,D_{{1}}\cos \,4\,\theta +
\,D_{{2}}\sin\,4\,\theta )
\Big) S'
 \nonumber \\
&+& 20\,\Big(
(\,B_{{2}}\cos \,2\,\theta - \,B_{{1}}\sin \,2\,\theta )
+
 14\, (D_{{2}}\cos \,4\,\theta - \,D_{{1}}\sin\, 4\,\theta )
\Big) S''
 \nonumber \\
 &+& 10\, \Big(
\,( B_{{1}}\cos \,2\,\theta  +  \,B_{{2}}\sin \,2\,\theta )
+ 4 (\,D_{{1}}\cos \,4\,\theta + \,D_{{2}}\sin \,4\,\theta)
 \Big) S^{(3)}
 \nonumber \\
 &+& \,\Big(
( B_{{1}}\sin \,2\,\theta  - \,B_{{2}}\cos \,2\,\theta )
+ 2 (\,D_{{1}}\sin \,4\,\theta- \,D_{{2}}\cos \,4\,\theta)
\Big) S^{(4)} \ ,
\label{LDES1}
\end{eqnarray}
\begin{eqnarray}
 0\ &=& \ 
 16\, \Big(
 (A_{{3}} \cos \,\theta + A_{{4}} \,\sin \,\theta)
-9(C_{{1}}\sin \,3\,\theta + \,C_{{2}}\cos \,3\,\theta)
\Big) S
 \nonumber \\
&+&
20\, \Big(
  (A_{{3}} \,\sin \, \theta - A_{{4}}\,\cos \,\theta)
+ 9\,( C_{{1}} \cos \,3\,\theta -  C_{{2}}\,\sin \,3\,\theta)
 \Big) S'
  \nonumber \\
 &+& 80\,\Big(\,C_{{1}}\sin \,3\,\theta + \,C_{{2}}\cos \,3\,\theta
\Big) S''
  \nonumber \\
 &+& 5\,\Big(
\,(A_{{3}} \sin \, \theta  - A_{{4}} \,\cos \,\theta)
- 3(C_{{1}}\,\cos \,3\,\theta - C_{{2}}\,\sin \,3\,\theta )
\Big)  S^{(3)}
  \nonumber \\
 &-& \,\Big(
  A_{{3}}\,\cos \,\theta     + A_{{4}}\,\sin \,\theta
+ C_{{1}}\,\sin \,3\,\theta  + C_{{2}}\,\cos \,3\,\theta
\Big) S^{(4)} \ .
\label{LDES2}
\end{eqnarray}

Let us now discuss these equations.
\begin{enumerate}
  \item The equation (\ref{gamma2}) determines $\gamma_2=\text{constant}$.
  \item Equations (\ref{eqa1}) and (\ref{eqb3}) must be used to determine $\gamma_1(\theta)$ and $\gamma_3(\theta)$ in terms of the function $S(\theta)$ figuring in the potential.
  \item Of the 12 ``doublet coefficients" figuring in the leading part of the integral $Y$, four figure only in (\ref{LDES1}), namely $B_1,\,B_2,\,D_1,\,D_2$ and four only in (\ref{LDES2}), namely $A_3,\,A_4,\,C_1,\,C_2$. The remaining ones ($A_1,\,A_2$) and ($B_3,\,B_4$) appear only in (\ref{eqa1}) and (\ref{eqb3}), respectively. By definition, the linear compatibility condition (\ref{CCsep}) must be satisfied trivially for exotic potentials at $R(r)=0$. Hence, for these potentials the only possible constants figuring in $Y$ are ($A_1,\,A_2$) and ($B_3,\,B_4$). This case was analyzed conclusively and in detail in our previous article\cite{ELW}.
\end{enumerate}

Let us now turn to the problem of finding all \emph{standard} superintegrable potentials. To do this we proceed as follows
\begin{itemize}
  \item I. Assume that at least one of the constants $B_1,\,B_2,\,D_1,\,D_2$ is not vanishing and solve (\ref{LDES1}) for $S(\theta)$.
  The general solution for $S(\theta)$ will depend on these constants and on additional integration constants.
  \item II. We proceed similarly with equation (\ref{LDES2}), i.e. solve it assuming that at least one of the $A_3,\,A_4,\,C_1,\,C_2$ is non-zero.

In both cases it is still necessary to solve the determining equations
\begin{eqnarray}
{\cal C}_r \ = \ 0 \ , \qquad {\cal C}_\theta \ = \ 0 \ ,
\end{eqnarray}
with ${\cal C}_r$ and ${\cal C}_\theta$ figuring in (\ref{commutator}), (see (\ref{Eq10p})-(\ref{Eq01p})). The determining equation  ${\cal C}_{r}=0$ defines $G_4(r,\theta)$, ${\cal C}_\theta \ = \ 0 $ will provide further information and
make it possible to determine $S(\theta)$ completely and hence also
the integrals $X$ and $Y$.
 \item III. It is also possible for both of the
 above sets of constants to contain nonzero constants.
 The corresponding potentials will then allow more than one
 fourth-order integral. However, at most $3$ integrals ($H$, $X$ and $Y$)
 can be polynomially independent. In any case, the function $S_{III}(\theta)$
 obtained in this case will be a special case of both $S_{I}$ and $S_{II}$ and 
 we shall not pursue this issue further. 
\end{itemize}

\subsection{General form of the angular part $S(\theta)$}
\textbf{Case I. $(B_1,\,B_2,\,D_1,\,D_2)\, \neq\,(0,\,0,\,0,\,0)$}
\vspace{0.2cm}

We concentrate in equation (\ref{LDES1}). We first define $T_I(\theta)$ by putting
\begin{equation}\label{TS}
  T'_I(\theta) \ \equiv \ S_I(\theta) \ .
\end{equation}

The general solution of the fifth-order linear ODE for $T_I$ obtained from (\ref{LDES1}) is
\begin{equation}
\label{TBDLinsol}
 T_{I} \left( \theta \right)\  = \  \frac {1}{M_I}({s_1}
+{s_2}\,\sin 2\,\theta
+{s_3}\,\cos 2\,\theta
+{s_4}\,\sin 4\,\theta
+{s_5}\,\cos 4\,\theta) \ ,
\end{equation}
where the $s_{i}$ are integration constants and
\[
M_I \ = \ B_{{1}}\sin   2\,\theta   - B_{{2}}\cos   2\,\theta
+2 (\,D_{{1}}\sin   4\,\theta    -  \,D_{{2}}\cos   4\,\theta) \ .
\]

Using (\ref{TS}) we obtain $S_I(\theta)$ as
\begin{equation}
 \begin{split}
 \label{SI}
S_{I}(\theta) \ = \ -\frac{1}{M_I^2}\bigg(  2\,s_1 [B_1\,\cos 2\,\theta +B_2\,\sin 2\,\theta + 4(D_1\,\cos 4\,\theta +D_2\,\sin 4\,\theta ) \,]
& \\
 + \ 2\,s_2\, [B_1 +D_1\,\cos 6\,\theta + D_2\,\sin 6\,\theta + 3(D_1\,\cos 2\,\theta +D_2\,\sin 2\,\theta ) \,]
& \\
+ \ 2\,s_3\, [B_2 -D_2\,\cos 6\,\theta + D_1\,\sin 6\,\theta + 3(D_2\,\cos 2\,\theta -D_1\,\sin 2\,\theta)\,]
& \\
+ \ s_4 \,[8\,D_1 -B_1\,\cos 6\,\theta - B_2\,\sin 6\,\theta + 3(B_1\,\cos 2\,\theta -B_2\,\sin 2\,\theta)\,]
& \\
+ \ s_5 \,[8\,D_2 + B_2\,\cos 6\,\theta - B_1\,\sin 6\,\theta + 3(B_2\,\cos 2\,\theta +B_1\,\sin 2\,\theta)\,]
   \bigg) \ .
 \end{split}
\end{equation}

There seem to be too many integration constants in (\ref{SI}), since (\ref{LDES1}) is a fourth-order ODE  for $S(\theta)$.
Indeed the five solutions corresponding to $s_1, \ldots, s_5$, are not independent and one can be eliminated in terms of the others. We shall not go into this here since constraints on the constants $s_i$ will be imposed by the last still unsolved determining equation, namely ${\cal C}_{\theta}=0$ and these will always leave less than 5 free constants.

\bigskip
\textbf{Case II $(A_3,\,A_4,\,C_1,\,C_2)\, \neq\,(0,\,0,\,0,\,0)$}
\bigskip

Similarly, the solution to the second linear equation (\ref{LDES2}) is
\begin{equation}
\label{TACLinsol2}
 T_{II} \left( \theta \right) = \frac {1}{M_{II}}(
 {s_1}
+{s_2}\,\sin \theta
+{s_3}\,\cos \theta
+{s_4}\,\sin 3\,\theta
+{s_5}\,\cos 3\,\theta) \ ,
\end{equation}
where
\[
M_{II} \ = \ A_{{3}} \cos   \theta   + A_{{4}} \sin   \theta
+C_{{1}}\sin   3\,\theta   + C_{{2}}\cos   3\,\theta \ .
\]
Consequently
\begin{equation}
 \begin{split}
 \label{SII}
S_{II}(\theta) \ = \ \frac{1}{M_{II}^2}\bigg(  2\,s_1 [A_3\,\sin\theta -A_4\,\cos \theta + 3(C_2\,\sin 3\,\theta -C_1\,\cos 3\,\theta ) \,]
& \\
 + \ s_2\, [A_3 -C_2\,\cos 4\,\theta - C_1\,\sin 4\,\theta + 2(C_2\,\cos 2\,\theta +C_1\,\sin 2\,\theta ) \,]
& \\
- \ s_3\, [A_4 +C_1\,\cos 4\,\theta - C_2\,\sin 4\,\theta + 2(C_1\,\cos 2\,\theta -C_2\,\sin 2\,\theta)\,]
& \\
+ \ s_4 \,[3\,C_2 +A_3\,\cos 4\,\theta + A_4\,\sin 4\,\theta + 2(A_3\,\cos 2\,\theta -A_4\,\sin 2\,\theta)\,]
& \\
- \ s_5 \,[3\,C_1 - A_4\,\cos 4\,\theta + A_3\,\sin 4\,\theta + 2(A_4\,\cos 2\,\theta +A_3\,\sin 2\,\theta)\,]
   \bigg) \ .
 \end{split}
\end{equation}
Again the five solutions corresponding to the $5$ constants $s_i$, are not linearly independent.

As mentioned above, the determining equation ${\cal C}_{r}=0$ (\ref{Eq10p}) defines the function $G_4(r,\,\theta)$
in both cases. The last equation ${\cal C}_{\theta}=0$ (\ref{Eq01p}) constrains the coefficients $s_ {i}$ in both 
cases. We again treat cases I and II, separately.

\clearpage

\textbf{Case I $(B_1,\,B_2,\,D_1,\,D_2)\, \neq\,(0,\,0,\,0,\,0)$}

\bigskip
In this case, from the equation ${\cal C}_{r}=0$ (\ref{Eq10p}) we obtain the function $G_4(r,\,\theta)$
\begin{equation}
\label{G40I}
\begin{aligned}
48\,r^4\,G_4(r,\theta) & \ = \   4\,\bigg(    32\, S\, S'\, \left(B_1 \sin 2\theta -B_2 \cos 2 \theta + 2 (D_1 \sin 4 \theta - D_2 \cos 4 \theta) \right)
\\ &  +  48\, S^2\, \left(B_1 \cos 2 \theta+B_2 \sin 2 \theta + D_1 \cos 4 \theta +D_2 \sin 4 \theta  \right)
\\ &   +  S'\, \big[S''\, \left(B_2 \cos 2 \theta - B_1\, \sin 2 \theta -2 (D_1 \sin 4 \theta - D_2 \cos 4 \theta) \right)
\\ & -6\, S'\, \left(B_1 \cos 2 \theta+B_2 \sin 2 \theta+ 4 (D_1 \cos 4 \theta  + D_2 \sin 4 \theta)  \right)\big]         \bigg)
\\ & + \hbar^2\,\bigg[  3072 \,S\,\left(D_1 \cos 4 \theta+D_2 \sin 4 \theta \right)
\\ &  + 64 \,S'\, \left(B_1 \sin 2 \theta -B_2 \cos 2 \theta +62 (D_1 \sin 4 \theta - D_2 \cos 4 \theta) \right)
\\ &  -96 \,S''\, \left(B_1 \cos 2 \theta+B_2 \sin 2 \theta  +20 \left(D_2 \sin 4 \theta +D_1 \cos 4 \theta \right)\right)
\\ &  + S^{(3)} \,\left(52 (B_2 \cos 2 \theta- B_1 \sin 2 \theta)  +440 \left(D_2 \cos 4 \theta -D_1 \sin 4 \theta \right)\right)
\\ &  + 12\, S^{(4)}\, \left(B_1 \cos 2 \theta+B_2 \sin 2 \theta  +4( D_2 \sin 4 \theta + D_1 \cos 4 \theta) \right)
\\ &  + S^{(5)} \,\left(B_1 \sin 2 \theta -B_2 \cos 2 \theta +2 (D_1 \sin 4 \theta - D_2 \cos 4 \theta ) \right)      \bigg] \ .
\end{aligned}
\end{equation}

\clearpage

The determining equation ${\cal C}_{\theta}=0$ reads
\begin{equation}\label{Ctheta}
\begin{aligned}
&  4\,\bigg(  256\, S^2\, \left(B_1 \sin 2 \theta -B_2 \cos 2 \theta +2 (D_1 \sin 4 \theta - D_2 \cos 4 \theta) \right)
\\ &  -240 \,S \,S' \,\big(B_2 \sin 2 \theta +B_1 \cos 2 \theta +4 (D_1 \cos 4 \theta+D_2 \sin 4 \theta)  \big)
\\ &  +60\, {(S')}^2 \,\left(B_2 \cos 2 \theta- B_1 \sin 2 \theta  -8 (D_1 \sin 4 \theta - D_2 \cos 4 \theta) \right)
\\ &  +40\, S\, S''\, \left(B_2 \cos 2 \theta- B_1 \sin 2 \theta -2 (D_1 \sin 4 \theta - D_2 \cos 4 \theta) \right)
\\ &  +30\, S' \,S''\,\left(B_1 \cos 2 \theta +B_2 \sin 2 \theta +4 (D_1 \cos 4 \theta+ D_2 \sin 4 \theta)  \right)
\\ &  + {(S'')}^2\, \left(B_1 \sin 2 \theta -B_2 \cos 2 \theta +2 (D_1 \sin 4 \theta - D_2 \cos 4 \theta) \right)
\\ &  + 3 \,S^{(3)}\, S'\, \big(B_1 \sin 2 \theta -B_2 \cos 2 \theta +2 (D_1 \sin 4 \theta - D_2 \cos 4 \theta) \big) \bigg)
\\ & + \hbar^2\,\bigg[  384\, S\, \left(B_2 \cos 2 \theta- B_1 \sin 2 \theta  -88 (D_1 \sin 4 \theta - D_2 \cos 4 \theta) \right)
\\ &  - 32\, S'\, \big(31( B_1 \cos 2 \theta+ B_2 \sin 2 \theta)  -1676 \left(D_1 \cos 4 \theta +D_2 \sin 4 \theta \right)\big)
\\ & +  16 \,S'' \,\big( 53 (B_2 \cos 2 \theta- B_1 \sin 2 \theta) +2114 \left(D_1 \sin 4 \theta -D_2 \cos 4 \theta \right)\big)
\\ &  -80\, S^{(3)} \,\left(B_1 \cos 2 \theta+B_2 \sin 2 \theta  +136 \left(D_1 \cos 4 \theta+D_2 \sin 4 \theta  \right)\right)
\\ &  -16 \,S^{(4)}\, \big(11 (B_1 \sin 2 \theta - B_2 \cos 2 \theta) +118 \left(D_1 \sin 4 \theta -D_2 \cos 4 \theta \right)\big)
\\ &  +  42\, S^{(5)}\, \left(B_1 \cos 2 \theta + B_2 \sin 2 \theta  +4 (D_1 \cos 4 \theta + D_2 \sin 4 \theta) \right)
\\ &  +  3 \,S^{(6)}\, \left(B_1 \sin 2 \theta -B_2 \cos 2 \theta +2 (D_1 \sin 4 \theta - D_2 \cos 4 \theta) \right) \bigg] \ = \ 0 \ .
\end{aligned}
\end{equation}

\clearpage

First, by an appropriate rotation we can always set $B_1=0$. Then, we substitute the function (\ref{SI}) in the remaining equation (\ref{Ctheta}). As a result we obtain algebraic equations for the $s_i$ ($i=1,2,3,4,5$) in (\ref{SI}) by setting the coefficients of $\cos k\theta$ and $\sin k\theta$ for $k=0,2,4,\ldots,18$ equal to zero. We obtain the following solutions
\begin{equation}\label{parI}
\begin{aligned}
& s_1^{(\ell)} \ = \ \frac{q_\ell}{4 \,D_1^2\,\hbar ^6\, \left(B_2^2-8 D_1^2\right){}^2 D_2^2 }\bigg[  B_2 D_2  \left(8 D_1^2-B_2^2\right) \left(B_2^2 \left(D_1^2+D_2^2\right)+8 D_1^2 \left(4 D_1^2+3 D_2^2\right)\right)\,\hbar ^6
\\ &
 \ + \  \left(8 B_2^2 \left(D_1^5-6 D_1^3 D_2^2\right)-B_2^4 \left(D_1^3+3 D_2^2 D_1\right)+64 \left(2 D_1^2+D_2^2\right) D_1^5\right)\,\hbar ^4\,q_\ell
\\ &
\- \ B_2 D_1^2 D_2 \left(3 B_2^2+40 D_1^2\right)\,\hbar ^2\,q_\ell^2 \ - \  D_1^3 \left(B_2^2+8 D_1^2\right)\,q_\ell^3         \bigg] \ ,
\\ & s_2^{(\ell)} \ = \ q_\ell\,\hbar^2 \ ,
&
\\ & s_3^{(\ell)}\ = \ \frac{q_\ell}{D_1\,D_2^2\,\hbar ^6\, \left(B_2^2-8 D_1^2\right){}^2 }\bigg[ D_2 \left(B_2^2-8 D_1^2\right) \left(3 B_2^2 \left(D_1^2+D_2^2\right)+8 D_1^2 \left(2 D_1^2+D_2^2\right)\right)\,\hbar ^6
\\ &
\ + \ \left(2 B_2^3 \left(D_1^3+4 D_2^2 D_1\right)-32 B_2 D_1^5\right)\,\hbar ^4\,q_\ell \ + \  D_1^2 D_2 \left(7 B_2^2+8 D_1^2\right)\,\hbar ^2\,q_\ell^2
\ + \ 2 \,B_2\, D_1^3\, q_\ell^3 \bigg]\ ,
\\ & s_4^{(\ell)} \ = \ \frac{q_\ell}{D_2^2\,\hbar ^6\, \left(B_2^2-8 D_1^2\right){}^2 }\bigg[ 4\,B_2 D_2 \left(D_1^2+2 D_2^2\right) \left(B_2^2-8 D_1^2\right)\,\hbar ^6
\\ &
\ + \ 2\,D_1 \left(B_2^2 \left(D_1^2+6 D_2^2\right)-16 D_1^2 \left(D_1^2+D_2^2\right)\right)\,\hbar ^4\,q_\ell
\ + \ 8 \,B_2\, D_1^2 \,D_2\,\hbar ^2\,q_\ell^2 \  + \ 2 \,D_1^3\, q_\ell^3   \bigg] \ ,
\\ & s_5 \ = \ 0 \ ,
\end{aligned}
\end{equation}
where $q_\ell,\,\ell=1,2,3,4$, are the four roots of the quartic equation
\begin{equation}\label{quartic}
\begin{aligned}
& D_1^4\, q^4 \ + \ 4 \,B_2\,D_2\, D_1^3\,\hbar ^2\,q^3 + (B_2^2 \left(D_1^4+6 D_2^2 D_1^2\right) \ - \  16 D_1^4 \left(D_1^2+D_2^2\right))\,\hbar ^4\,q^2
\\ & \ - \ 2\,B_2\, D_1\, \left(8 D_1^2-B_2^2\right) D_2 \left(D_1^2+2 D_2^2\right)\,\hbar ^6 \,q
\ + \  \left(B_2^2-8 D_1^2\right){}^2 D_2^2 \left(D_1^2+D_2^2\right)\,\hbar ^8 \ = \ 0 \ .
\end{aligned}
\end{equation}

The discriminant $\Gamma$ of the quartic equation (\ref{quartic}) is given by
\begin{equation}\label{disqu}
\Gamma \ = \ -256\,\hbar ^{24}\, D_1^{24} \,D_2^2\, \left(B_2^2-8 D_1^2\right){}^2 \left[B_2^4 \left(60 D_2^2-48 D_1^2\right)+768 B_2^2 \left(D_1^2+D_2^2\right){}^2+B_2^6-4096 \left(D_1^2+D_2^2\right){}^3\right] \ .
\end{equation}
This discriminant is zero if and only if at least two roots of (\ref{quartic}) are equal. If the discriminant is negative there are two real roots and two complex conjugate roots. If it is positive the roots are either all real or all non-real. From a physical point of view only the real solutions are admitted. Let us analyze the zeros of the discriminant (\ref{disqu}). They will correspond to the TTW model\cite{TTW:2009, TTW:2010}. For $\Gamma \neq 0$, by substituting (\ref{parI}) into (\ref{SI}) we obtain an angular component $S_I(\theta)$ proportional to $\hbar^2$ with no classical analog, it cannot be transformed or reduced to that of the TTW model.

\subsubsection{Case $\hbar=0$}
For $\hbar=0$, the discriminant (\ref{disqu}) vanishes. The corresponding coefficients take the values
\[
s_1 = s_1\ , \qquad s_2 = 0\ , \qquad s_3 = 0\ , \qquad s_4 = s_4\ , \qquad s_5 = s_5 \ ,
\]
which yields the potential
\begin{equation}
\begin{split}
S_I(\theta) &=
\label{}
\frac {4\,( D_{{1}} \cos  4\,\theta   + \,D_{{2}} \sin  4\,\theta  ) s_{{1}} +  4\,(D_{{1}}s_{{5}}+\,D_{{2}}s
_{{4}})}{ \left( {D_{{1}}}^{2}-{D_{{2}}}^{2} \right)
\cos \,8\,\theta +2\,D_{{1}}D_{{2}}\sin \, 8\,\theta
- ( {D_{{1}}}^{2}+{D_{{2}}}^{2})}\,.
\end{split}
\end{equation}
For $D_{{2}}=0$, the angular part of the potential becomes
\begin{equation}\label{TTWSI}
 S_I(\theta) = \frac {  {\tilde s_1}\,\cos \, 4\,\theta + {\tilde s_{{5}}}  }
{\cos \,8\,\theta - 1}\,,
\end{equation}
where $ {\tilde s_1} = 4\,s_1/D_1$, $ {\tilde s_5} = 4\,s_5/D_1$. This
potential (\ref{TTWSI}) coincides with that of the TTW Hamiltonian\cite{TTW:2009, TTW:2010}. The angular component of the TTW potential is
\begin{equation}
\begin{split}
\label{TTW}
S_{\rm TTW} (\theta) &={\frac {\alpha\,{k}^{2}}{  \cos^2 \left( k\theta \right)  ^{
}}}+{\frac {\beta\,{k}^{2}}{  \sin^2 \left( k\theta 
 \right) ^{}}}
 \\&=
 {\frac {4\,{k}^{2}\, (\alpha\,-\, \beta)\cos  2\,k\,\theta\ - \ 4\,{k}^{2}(\alpha\,+\,\beta\,)}{\cos  4\,k\,\theta  \ - \ 1}}\ .
 \end{split}
\end{equation}

Then for $k=2$ and \(  \alpha=  ({\tilde s_1} - {\tilde s_5})/32  \,, \beta=
-({\tilde s_{{1}}} + {\tilde s_{{5}}})/32\), the potential (\ref{TTWSI}) actually corresponds to a rotated TTW model (with no radial component $R(r)=0$) which is a superintegrable system in both the classical and quantum cases.

\subsubsection{Case $D_1=0$}
For $D_1=0$, the discriminant (\ref{disqu}) vanishes as well. The corresponding coefficients vanish, $s_1=s_2=s_3=s_4=s_5=0$, which gives the trivial solution $S_I(\theta)=0$.

\subsubsection{Case $D_2=0$}
For $D_2=0$, the discriminant (\ref{disqu}) vanishes again. The corresponding coefficients are given by
\[
s_1 = s_1 \ , \qquad s_2 = 0 \ , \qquad s_3 = \frac{B_2^2\,s_4-8\,D_1^2\,(s_1+s_4)}{2\,B_2\,D_1}\ , \qquad s_4=s_4 \ , \qquad s_5 = 0  \ ,
\]
thus
\[
S(\theta) \ = \ -\frac{2(B_2\,s_4 + 2\,D_1\,s_1\,\sin 2\,\theta +2\,D_1\,s_4\,\sin 2\,\theta    )}{B_2\,D_1\,(1+\cos4\,\theta)} \ .
\]

This solution corresponds to the angular component of the TTW model (\ref{TTW}) with $k=1$. The zeros of the discriminant (\ref{disqu}) correspond to a TTW model.

\bigskip
\textbf{Case II: $(A_3,\,A_4,\,C_1,\,C_2)\, \neq\,(0,\,0,\,0,\,0)$}
\vspace{0.4cm}

Now, we proceed to the case II. First, from the determining equation ${\cal C}_{r}=0$ (\ref{Eq10p}) we obtain the function $G_4(r,\,\theta)$
{
\begin{equation}
\label{G40II}
\begin{aligned}
48\,r^3\,G_4(r,\theta) & \ = \ 16 \,S'\,\bigg[  3 \,S'\, \left(A_4 \cos \theta- A_3 \sin \theta  -3 (C_2 \sin 3 \theta - C_1 \cos 3 \theta) \right)
\\ &  + \ (S''- 14\, S\,) \big(A_3 \cos \theta+A_4 \sin \theta  +C_1 \sin 3 \theta +C_2 \cos 3 \theta \big)      \bigg]
\\ &  + \hbar^2\,\bigg[  16\, S\, \big(2 (A_4 \cos \theta - A_3 \sin \theta) +15( C_2 \sin 3 \theta - C_1 \cos 3 \theta) \big)
\\ &  4\, S'\, \big(7 (A_3 \cos \theta+ A_4 \sin \theta)  -127 \left(C_1 \sin 3 \theta +C_2 \cos 3 \theta \right)\big)
\\ &  +  4 \,S''\, \big(A_3 \sin \theta -A_4 \cos \theta + 93 (C_1 \cos 3 \theta- C_2 \sin 3 \theta)  \big)
\\ &  +  S^{(3)}\, \big(9 (A_3 \cos \theta + A_4 \sin \theta) +121 \left(C_1 \sin 3 \theta +C_2 \cos 3 \theta \right)\big )
\\ &  +  6 \,S^{(4)}\, \big(A_3 \sin \theta -A_4 \cos \theta +3 (C_2 \sin 3 \theta - C_1 \cos 3 \theta ) \big)
\\ &  -  S^{(5)}\, (A_3 \cos \theta +  A_4 \sin \theta  +C_1 \sin 3 \theta +C_2 \cos 3 \theta )     \bigg] \ ,
\end{aligned}
\end{equation}
}

\clearpage
and correspondingly ${\cal C}_{\theta}=0$ (\ref{Eq01p}) takes the form
{
\begin{equation}\label{Ctheta2}
\begin{aligned}
&  4\,\bigg[  -36 \,S^2\, \left(A_3 \cos \theta + A_4 \sin \theta +C_1 \sin 3 \theta +C_2 \cos 3 \theta \right)
\\ &  +  15 \,S'^2 \,\big(A_3 \cos \theta+A_4 \sin \theta  +9( C_1 \sin 3 \theta + C_2 \cos 3 \theta) \big)
\\ &   +  20\, S\, S''\, \left(A_3 \cos \theta + A_4 \sin \theta +C_1 \sin 3 \theta +C_2 \cos 3 \theta \right)
\\ &  -  {(S'')}^2\, \left(A_3 \cos \theta + A_4 \sin \theta  +C_1 \sin 3 \theta +C_2 \cos 3 \theta \right)
\\ & +3\,S'\,\bigg(  20 \,S\, \big(A_4 \cos \theta -A_3 \sin \theta +3 (C_1 \cos 3 \theta - C_2 \sin 3 \theta )\big)
\\ &  + 5\, S''\, \big(A_3 \sin \theta -A_4 \cos \theta +3 (C_2 \sin 3 \theta - C_1 \cos 3 \theta )\big)
\\ &  -  S^{(3)}\, \left(A_3 \cos \theta + A_4 \sin \theta  +C_1 \sin 3 \theta +C_2 \cos 3 \theta \right)   \bigg) \bigg]
\\ & + \hbar^2\,\bigg[ 96 \,S\, \big(A_3 \cos \theta + A_4 \sin \theta  +39 \left(C_1 \sin 3 \theta +C_2 \cos 3 \theta \right)\big)
\\ &  +  24\, S'\,\big(A_3 \sin \theta -A_4 \cos \theta +303 \left(C_2 \sin 3 \theta -C_1 \cos 3 \theta \right)\big)
\\ &  +   8\, S''\, \big(21( A_3 \cos \theta + A_4 \sin \theta)  -719 \left(C_1 \sin 3 \theta +C_2 \cos 3 \theta \right)\big)
\\ &   +  30 \,S^{(3)}\, \big(3 (A_3 \sin \theta - A_4 \cos \theta )+79 (C_1 \cos 3 \theta -  C_2 \sin 3 \theta ) \big)
\\ &  +  8\, S^{(4)}\, \big( 3 (A_3 \cos \theta +A_4 \sin \theta )+67 \left(C_1 \sin 3 \theta +C_2 \cos 3 \theta \right)\big)
\\ &  +  21 \, S^{(5)} \, \big(A_3 \sin \theta - A_4 \cos \theta  + 3( C_2 \sin 3 \theta - C_1 \cos 3 \theta) \big)
\\ &  - 3\, S^{(6)}\, \left(A_3 \cos \theta + A_4 \sin \theta +C_1 \sin 3 \theta +C_2 \cos 3 \theta \right) \bigg] \ = \ 0   \ .
\end{aligned}
\end{equation}
}

\clearpage
Now we focus on the potential $S_{II}(\theta)$ (\ref{SII}).  By an appropriate rotation we can always set $A_3=0$. Then, we substitute the expression (\ref{SII}) into the remaining determining equation (\ref{Ctheta}). As a result we obtain algebraic equations for the coefficients $s_i$ ($i=1,2,3,4,5$) by setting the coefficients of $\cos k\theta$ and $\sin k\theta$ for $k=1,2,3,\ldots,16$ equal to zero. In general, we obtain the following three solutions
\begin{equation}\label{parII}
\begin{aligned}
& s_1 \ = \ 0 \ ,
\\ & s_2 \ = \ s_2 \ ,
\\ & s_3^{(\ell)} \ = \ q_\ell \ ,
\\ & s_4^{(\ell)} \ = \  \frac{\hbar ^4 \left(\left(2 C_1^2+C_1A_4+3 C_2^2\right) q_\ell+C_1 C_2 s_2\right) -\left(C_1+A_4\right) \hbar ^2 q_\ell^2 -q_\ell^3}{A_4\,C_2\, \hbar ^4} \ ,
\\ &  s_5^{(\ell)} \ = \  \frac{ \hbar ^4 \left(\left(4 C_1^2+C_1A_4+6 C_2^2-A_4^2\right) q_\ell+\left(2 C_1+A_4\right) C_2 s_2\right) -\left(2 C_1+3A_4\right) \hbar ^2 q_\ell^2-2 q_\ell^3   }{A_4\,\left(2 C_1+A_4\right) \hbar ^4} \ ,
\end{aligned}
\end{equation}
where $q_\ell,\, \ell=1,2,3$, is the solution to the cubic equation
\begin{equation}\label{cubic}
\begin{aligned}
& q^3+ A_4\hbar ^2\,q^2 - \hbar ^4\,q\, \left(3 C_1^2 +2 A_4C_1 +3 C_2^2 \right)+\left(2 C_1^3+A_4C_1^2+2 C_2^2 C_1+A_4C_2^2\right) \hbar ^6\ = \ 0 \ .
\end{aligned}
\end{equation}

Again, substituting (\ref{parII}) into (\ref{SII}) we obtain the corresponding angular component $S_{II}(\theta)$ of the potential. Notice that there exist potentials $S_{II}(\theta)$ proportional to $\hbar^2$ that have no classical analog.

The discriminant $\Omega$ of the cubic equation (\ref{cubic}) is given by
\begin{equation}\label{discu}
\Omega \ =\ -4\,\hbar^{12}\,C_2^2\,\big[ A_4^4 \ + \ 8\,A_4^3\,C_1 \ + \ 18\,A_4^2\,(C_1^2+C_2^2) \ - \ 27\,{(C_1^2+C_2^2)}^2     \big]   \ .
\end{equation}

The discriminant (\ref{discu}) is zero if and only if at least two roots are equal. It is positive if the roots are all distinct real numbers, and negative if there exist one real and two complex conjugate roots.

For $\hbar=0$, the discriminant (\ref{discu}) vanishes. The corresponding coefficients take the values
\[
s_1 = s_1\ , \qquad s_2 = 0\ , \qquad s_3 = 0\ , \qquad s_4 = s_4\ , \qquad s_5 = s_5 \ ,
\]
which yields the potential
\begin{equation}
\begin{split}
S_{II}(\theta) \ & = \
\label{}
-\frac {6\,(s_5 + s_1\,\cos 3\,\theta)}{C_1\,(1- \cos 6\,\theta)}\ ,
\end{split}
\end{equation}
($A_4=C_2=0$) it corresponds to that of the TTW potential with $k=\frac{3}{2}$. Similarly we can show that for $C_2=0$, thus $\Omega=0$, the TTW (angular) potentials with $k=\frac{1}{2}$ and $k=1$ occur.

\bigskip
\section{CONFINING POTENTIALS}
\label{CPotentials}

\subsection*{Deformed Coulomb Potential}
Here we address the case of the deformed Coulomb potential
\begin{equation}\label{CPoKep}
  V(r,\,\theta) \ = \ \frac{a}{r} \ + \ \frac{S(\theta)}{r^2} \ ,
\end{equation}
where $a \neq 0$ is a real constant.

Similarly to the non confining case $R(r)=0$, from (\ref{Eq30p}), (\ref{Eq21p}) and (\ref{Eq12p}) corresponding
to the determining equations ${\cal A}_{rrr}={\cal A}_{rr\theta}={\cal
A}_{r\theta\theta}=0$, respectively, we determine the functions
$G_{1}(r,\theta)$, $G_{3}(r,\theta)$, and $G_{2}(r,\theta)$ up to arbitrary additive functions $\gamma_1(\theta)$, $\gamma_3(\theta)$ and $\gamma_2(\theta)$, respectively.

Then, we substitute such functions $G_{1,2,3}$ into the determining equation ${\cal A}_{\theta\theta\theta}=0$ (\ref{Eq03p}). By doing so, such equation yields again a set of five linear ODE
\begin{itemize}
  \item 1. Two fourth-order ODE for the angular component $S(\theta)$. The first ODE contains
the doublets $(B_1,B_2),(D_1,D_2)$, while the second ones depends on the doublets $(A_3,A_4),(C_1,C_2)$ and the parameter $a$ figuring in the Coulomb term.
  \item 2. Three linear ODE that define the functions $\gamma_1(\theta)$, $\gamma_2(\theta)$ and $\gamma_3(\theta)$, respectively. The equation for $\gamma_1$ depends on the doublet $(B_3,B_4)$ and the parameter $a$. For the function $\gamma_3$ the doublet $(A_1,A_2)$ and the parameter $a$ are involved and the equation for $\gamma_2$ contains only the parameter $a$.
\end{itemize}

For $a \neq 0$ the most general form for the angular component $S(\theta)$ is given by the Case II (\ref{SII}) for which at least one of the constants $C_1,\,C_2,\,A_3,\,A_4$ is not vanishing. Therefore, the function $S(\theta)$ is not new with respect to the non-confining case $R(r)=0$ and it was already described in detail in the Section \ref{Potentials}.

The corresponding functions $G_{1,2,3}$ are given by
\begin{equation}
\label{}
\begin{aligned}
& G_1(r,\theta) \ = \ G_1^{(0)}(r,\theta) \ ,
\\ & G_2(r,\theta) \ = \ G_2^{(0)}(r,\theta)  \ + \   5\,a\,\bigg[    \frac{A_4 \cos \theta - A_3 \sin  \theta +3 (\,C_1 \cos 3 \theta -\, C_2 \sin 3 \theta) }{2\, r^2} \bigg] \ ,
\\ & G_3(r,\theta) \ = \ G_3^{(0)}(r,\theta)  \ + \   3\,a\,\bigg[    \frac{A_3 \cos \theta + A_4 \sin  \theta +\,C_2 \cos 3 \theta + C_1\, \sin 3 \theta }{ r} \bigg] \ ,
\end{aligned}
\end{equation}
where the $G_i^{(0)}(r,\theta)$, $i=1,2,3$, coincide with those of the non-confining potential given by (\ref{G1gen}), (\ref{G3gen}) and (\ref{G2gen}), respectively, putting $B_1=B_2=D_1=D_2=0$ and $\gamma_1=\gamma_2=\gamma_3=0$.

From the determining equation ${\cal C}_{r}=0$ given in (\ref{Eq10p}) we calculate the corresponding function $G_4(r,\theta)$
\begin{equation}
\begin{aligned}
& G_4(r,\theta) \ = \  G_4^{(0)}(r,\theta) \ + a\,\bigg[\
 \frac{1}{r^2}\bigg( \frac{1}{4} \hbar ^2 \left(7 (A_4 \cos \theta- A_3 \sin \theta)  +99 (C_2 \sin 3 \theta - C_1 \cos 3 \theta) \right)
\\ &\qquad \qquad +5 S'(\theta ) \left(A_4 \sin \theta +A_3 \cos \theta +C_1 \sin 3 \theta +C_2 \cos 3 \theta \right) \bigg) \ ,
\end{aligned}
\end{equation}
here $G_4^{(0)}(r,\theta)$ is given in (\ref{G40II}). The angular component $S(\theta)$ is given by (\ref{SII}) with the coefficients $s_i$ (\ref{parII}) presented in the Case II.

\subsection*{Deformed Harmonic Oscillator Potential}
Here we address the case of the confining potential
\begin{equation}\label{CPoHar}
  V(r,\,\theta) \ = \ b\,r^2 \ + \ \frac{S(\theta)}{r^2} \ ,
\end{equation}
with $b \neq 0$ a real constant.

Following the same strategy as for the previous case, we obtain the following functions
\begin{equation}
\label{}
\begin{aligned}
& G_1(r,\theta) \ = \ G_1^{(0)}(r,\theta)\ + \ 2\,b\,r^2\,(B_1 \cos 2 \theta +B_2 \sin 2 \theta +D_1 \cos 4 \theta +D_2\sin 4 \theta ) \ ,
\\ & G_2(r,\theta) \ = \ G_2^{(0)}(r,\theta) \ ,
\\ & G_3(r,\theta) \ = \ G_3^{(0)}(r,\theta) \ + \ 2\,b\,r\,(B_2 \cos  2\theta - B_1 \sin 2\theta + 2\,(D_2 \cos  4\theta - \,D_1 \sin 4\theta))\ ,
\\ & G_4(r,\theta) \ = \  G_4^{(0)}(r,\theta)  \ ,
\end{aligned}
\end{equation}
where the $G_i^{(0)}(r,\theta)$, $i=1,2,3$, coincide with those of the non-confining potential $R(r)=0$, given in (\ref{G1gen}), (\ref{G3gen}) and (\ref{G2gen}), respectively, putting $A_3=A_4=C_1=C_2=0$ and $\gamma_1=\gamma_2=\gamma_3=0$. The function $G_4^{(0)}(r,\theta)$ is given in (\ref{G40I}).

For $b\neq0$, the angular component $S(\theta)$ is given by (\ref{SI}) with the coefficients $s_i$ (\ref{parI}) presented in the Case I for which at least one of the constants $B_1,\,B_2,\,D_1,\,D_2$ is not vanishing.

\section{CONCLUSIONS}
\label{Conclusions}
We considered superintegrable systems in a two-dimensional Euclidean space. Classical
and quantum fourth-order superintegrable \emph{standard} potentials separating in polar
coordinates were derived. We can summarize the main results via the following
theorems

\vspace{0.3cm}

{\bf Theorem 1.} \emph{In classical mechanics, the standard superintegrable confining systems, $a\,b\neq 0$,
correspond to the TTW potential}
\begin{eqnarray}
V_{\rm TTW}(r,\,\theta) \ = \  b\,r^2 \  + \  \frac{1}{r^2}\bigg[ \frac{\alpha}{\cos^2 2\,\theta}  \ + \ \frac{\beta}{\sin^2 2\,\theta}   \bigg] \ ,
\end{eqnarray}
($\alpha,\,\beta$ \emph{real constants}), \emph{and the PW potential}
\begin{eqnarray}
V_{\rm PW}(r,\,\theta) \ = \ \frac{a}{r} \  + \  \frac{1}{r^2}\bigg[ \frac{\mu}{\cos^2 \frac{3}{2}\,\theta}  \ + \ \frac{\nu}{ \sin^2 \frac{3}{2}\theta}   \bigg] \ ,
\end{eqnarray}
($\mu,\,\nu$ \emph{real constants}). \emph{The corresponding leading terms of the integral Y in}
(\ref{Y}) \emph{are proportional to}
\[
  (p_x^4+p_y^4-6\,p_x^2\,p_y^2)   \ ,
\]
\emph{and}
\[
\{L_z,\ 3\,p_x^2\,p_y-p_y^3 \} \ ,
\]
\emph{respectively. These terms are independent of $a$ and $b$.}

\bigskip

{\bf Theorem 2.} \emph{In quantum mechanics, the new confining superintegrable systems correspond to }
\[
V(r,\,\theta) \ = \ b\,r^2 \ + \ \frac{1}{r^2}\,S_{I}(\theta) \ ,
\]
\emph{and}
\[
V(r,\,\theta) \ = \ \frac{a}{r} \ + \ \frac{1}{r^2}\,S_{II}(\theta)\ ,
\]
\emph{where} $S_I(\theta)$ \emph{is given by}
(\ref{SI}) with (\ref{parI}), \emph{and} $S_{II}(\theta)$ \emph{takes the form} (\ref{SII}) with (\ref{parII}). \emph{In general, the functions} $S_I$ \emph{and} $S_{II}$ \emph{are proportional to} $\hbar^2$ \emph{and cannot be reduced to a TTW or PW potential.} \emph{The corresponding leading terms of the integral Y in} (\ref{Y}) \emph{are}
\[
 2\,B_2\,p_x\,p_y\,(p_x^2+p_y^2)  \ + \  D_1\,(p_x^4+p_y^4-6\,p_x^2\,p_y^2)  \ + \  4\,D_2\,p_x\,p_y\,(p_x^2-p_y^2) \ ,
\]
$(B_2\,,D_1\,,D_2) \neq (0,0,0)$ \emph{and}
\[
A_4\,\{L_z,\ p_y\,(p_x^2+p_y^2)    \} \  + \  C_1\,\{L_z,\ 3\,p_x^2\,p_y-p_y^3 \} \  +  \ C_2\,\{L_z,\ p_x^3 -3\,p_y^2\,p_x \} \ ,
\]
$(A_4\,,C_1\,,C_2) \neq (0,0,0)$ \emph{respectively.}

\emph{We emphasize that these are pure quantum potentials. As particular cases, both potentials $V_{\rm TTW}$ and $V_{\rm PW}$ appear in the quantum case as well.}


Work is currently in progress on a general and unified 
study of $N^{th}$-order \emph{exotic} and \emph{standard} potentials separating in polar coordinates\cite{EWY}. Within this study\cite{EWY} the TTW and PW models correspond to \emph{standard} classical potentials.
For some cases we plan to present the polynomial algebra generated by the
integrals of motion and to use it to calculate the energy spectrum and the wave
functions in the quantum case.

\section{ACKNOWLEDGMENTS}
The research of PW was partially supported by a research grant from NSERC of
Canada. JCLV thanks PASPA grant (UNAM, Mexico) and the Centre de Recherches
Math\'ematiques, Universit\'e de Montr\'eal for the kind hospitality while on
sabbatical leave during which this work was done. The research of AME was
partially supported by a fellowship awarded by the Laboratory of Mathematical Physics of the CRM.
The research of \.{I}Y was partly supported by Hacettepe University Scientific Research
Coordination Unit. Project Number: FBI-2017-14035. He also thanks the Centre de
Recherches Math\'{e}matiques, Universit\'{e} de Montr\'{e}al and especially Professor
Pavel Winternitz for kind hospitality during his sabbatical leave.

\newpage
\section*{Appendix}
Explicitly, the functions $F_{1}, \ldots ,F_{13}$ figuring in (\ref{Eq30p})-(\ref{Eq01p}) are given by
\begin{eqnarray}
F_1 &=& 2 \Big(
\,  { B}_{{1}}\,\cos 2\,\theta
+\, { B}_{{2}}\,\sin  2\,\theta
+\, { D}_{{1}}\,\cos  4\,\theta
+\, { D}_{{2}}\,\sin  4\,\theta
\Big)\,,
\nonumber
 \\ 
F_2 &=&
  \frac{1}{r}
\Big({B}_{{2}}\,\cos 2\,\theta
-    {B}_{{1}}\,\sin 2\,\theta
- 2\,({D}_{{1}}\,\sin 4\,\theta
-\,{D}_{{2}}\,\cos 4\,\theta)
\Big)  \nonumber \\ 
&+&     {A}_{{3}}\,\cos\, \theta
+              {A}_{{4}}\,\sin\, \theta
+   {C}_{1}\,  \sin 3\,\theta
+              {C}_{2}\,  \cos 3\,\theta
\,,
\nonumber \\ 
F_3 &=&
 \frac{1}{ {r}^{3} } \Big(
   {B}_{{2}}\,\cos  2\theta
-  {B}_{{1}}\,\sin  2\theta
+ 2({D}_{{1}}\,\sin  4\theta
- {D}_{{2}}\,\cos  4\theta)
\Big)
\nonumber
\\ 
&+& \frac{1}{{r}^{2}}
\Big(
  {A}_{{4}}\,\sin \theta
+ {A}_{{3}}\,\cos \theta
- 3\,( {C}_{2} \cos  3\,\theta
+ {C}_{1} \sin  3\,\theta)
\Big)
\nonumber
\\ 
&-& \frac{2}{{r}}
\Big(
{B}_{{3}} \,\sin 2\,\theta
-{B}_{{4}}\,\cos 2\,\theta
\Big)
+ {A}_{{2}}\,\sin \theta + {A}_{{1}}\,\cos \theta
\,,
\nonumber \\ 
F_4 &=&
\frac{2}{{r}^{4}}
\Big(
  \, {D}_{{1}}\,\cos 4\,\theta
+ \, {D}_{{2}}\,\sin 4\,\theta
- \, {B}_{{1}}\,\cos 2\,\theta
- \, {B}_{{2}}\,\sin 2\,\theta
\Big)
\nonumber \\ 
&+& \frac{4}{{r}^{3}}
\Big(
  {A}_{{4}}\,\cos \theta
- {A}_{{3}}\,\sin \theta
- {C}_{1}\, \cos 3\,\theta
+ {C}_{2}\, \sin 3\,\theta
\Big)
\nonumber \\ 
&-& \frac{4}{{r}^{2}}
\Big(
  {B}_{{3}}\,  \cos  2\,\theta
+ {B}_{{4}} \,\sin  2\,\theta
\Big)
+ \frac{4}{{r}}
\Big(
  {A}_{{2}}  \,\cos \theta
- {A}_{{1}} \,\sin \theta
\Big)\,,
\nonumber \\  
F_5 &=&
- \frac{6}{{r}^{2}}
\Big(
   \,{ D}_{{1}}\,\cos  4\,\theta
+  \,{ D}_{{2}}\,\sin  4\,\theta
\Big)
+ \frac{2}{{r}}
\Big(
    {A}_{{4}} \,\cos \theta
-   {A}_{{3}} \,\sin \theta
\nonumber \\
&+& 3(\,{C}_{1} \cos  3\,\theta
- {C}_{2} \sin  3\,\theta)
\Big)
+ 2\,({B}_{{3}} \,\cos  2\,\theta
+ {B}_{{4}} \,\sin  2\,\theta)
\,,
\nonumber \\
F_6 &=&
\frac{3}{2}
\left(
  A_4 \,{\cos \theta }
- A_3 \,{\sin \theta }
+ 3 (C_1 \,{\cos 3\theta}
-    C_2 \,{\sin 3\theta })\right)
-\frac{9}{r} \left(
   D_1 \,{\cos 4\,\theta }
+  D_2 \,{\sin 4\,\theta }\right)
\,,
\nonumber \\
F_7 &=&
\frac{12}{r^2}
\left(
  \,D_1 \,{\sin 4\theta }
- \,D_2 \,{\cos 4\theta }
\right)
 +  \frac{1}{2 \,r} \left(
  A_4 \,{\sin \theta }
+  A_3 \,{\cos \theta }
\right. \nonumber \\  
&-& \left. 15 (\,C_1 \,{\sin 3\theta }
+ \,C_2 \,{\cos 3\theta } ) \right)
-  2\,(B_3 \,{\sin 2\theta}
-B_4 \,{\cos 2\theta})
\,,
\nonumber \\
F_8 &=& -\frac{3}{r^3} \left(
   \,B_1 \,{\cos 2\theta }
+  \,B_2 \,{\sin 2\theta }
-5( \,D_1 \,{\cos 4\theta }
+ \,D_2 \,{\sin 4\theta })
\right) \nonumber \\ 
&+& \frac{3}{2 r^2} \left(
   A_4 \,{\cos \theta }
-  A_3 \,{\sin \theta }
- 9(\,C_1 \,{\cos 3\theta }
-\,C_2 \,{\sin 3\theta })
\right) \nonumber \\ 
&-& \frac{5}{r}  \left(
  B_3 \,{\cos 2\theta}
+ B_4 \,{\sin 2\theta}
\right)
+\frac{3}{2} \left(
   A_2 \, {\cos \theta }
-  A_1 \, {\sin \theta } \right)
\,,
\nonumber \\
F_9 &=& \frac{9}{r^4} \left(
   B_1 \,{\sin 2\theta }
-  B_2 \,{\cos 2\theta }
-2 \,(D_1 \,{\sin 4\theta }
- \,D_2 \,{\cos 4\theta })
\right) \nonumber \\ 
&+& \frac{15}{2 r^3} \left(
3\,(C_1 \, {\sin 3\theta }
+C_2 \, {\cos 3\theta })
- A_3 \,{\cos \theta }
- A_4 \,{\sin \theta }
\right) \nonumber \\ 
&+& \frac{12}{r^2} \left(
   B_3 \,{\sin 2\theta }
-  B_4 \,{\cos 2\theta }
\right)
 -\frac{9}{2 \,r} \left(
 A_1 \, {\cos \theta }
 + A_2 \, {\sin \theta }
\right)
\,,
\nonumber \\
F_{10} &=& \frac{3}{r} \,\left(
\,({B_2}\,{\cos 2\,\theta}
- \,{B_1}\,{\sin 2\,\theta}) 
+2\,({D_2}\,{\cos 4\theta }
- \,{D_1}\,{\sin 4\theta})\right) \nonumber \\ 
&+& 3\,\big({A_3}\,{\cos \theta}
+ {A_4}\,{\sin \theta} 
+ C_1\, {\sin 3\,\theta}
+ C_2\, {\cos 3\,\theta}))
\,,
\nonumber \\
F_{11} &=& -\frac{3}{r^2} \left(
    B_1\, {\cos 2\theta }
+   B_2 \,{\sin 2\theta }
- 5 \,(D_1 \,{\cos 4\theta }
+\, D_2 \,{\sin 4\theta })
\right) - 4\, \left(
  B_3 \,  {\cos 2\theta }
+ B_4 \,  {\sin 2\theta }
\right) \nonumber \\ 
&+& \frac{1}{r} \left(
   A_4 \, {\cos \theta }
 - A_3 \, {\sin \theta }
- 9\,(C_1 \,  {\cos 3\theta }
- \,C_2 \,  {\sin 3\theta }) \right)
\,,
\nonumber 
\end{eqnarray}
\begin{eqnarray}
F_{12} &=& 
\frac{2}{r^3}  \left(
  B_1 \,{\sin 2\theta }
- B_2 \,{\cos 2\theta }
- 14 \,(D_1 \,{\sin 4\theta }
-\,D_2 \,{\cos 4\theta })
\right) \nonumber \\ 
&+& \frac{3}{2\, r^2}   \left(
 11\,(C_1\, {\sin 3\theta }
+ \,C_2\, {\cos 3\theta })
-A_3 \,{\cos \theta }
- A_4 \,{\sin \theta }
\right) \nonumber \\ 
&+& \frac{6}{r}  \left(
  B_3 \,{\sin 2\theta }
- B_4 \,{\cos 2\theta }
\right) - \frac{3}{2} \left(
  A_1 \,{\cos \theta }
+ A_2 \,{\sin \theta }
 \right)
\,,
\nonumber \\  
F_{13} &=& \frac{4}{r^4}   \left(
 2\,( B_1\, {\cos 2\theta }
+ \,B_2 \,{\sin 2\theta })
-11( \,D_1 \,{\cos 4\theta }
+ \,D_2 \,{\sin 4\theta })
\right) \nonumber \\ 
&-& \frac{2}{r^3}
\left(
  A_4  \,  {\cos \theta }
- A_3  \,  {\sin \theta }
- 17\,(C_1 \,  {\cos 3\theta}
-\,C_2 \,  {\sin 3\theta})
\right) \nonumber \\ 
&+& \frac{12}{r^2} \left(
   B_3\, {\cos 2\theta }
+  B_4\, {\sin 2\theta } \right)
- \frac{3}{r} \left(
   A_2 \, {\cos \theta }
-  A_1 \, {\sin \theta }
\right)
\,.
\nonumber
\end{eqnarray}

\end{document}